\documentstyle[aps]{revtex}
\begin{document}
\sloppy

\title{ 
Comment on ``Mixing and Decay Constants of Pseudoscalar Mesons''}
\author{M. Kirchbach \\
{\small \it Escuela de Fisica, Univ. Aut. de Zacatecas,\\
Apartado Postal C-580, Zacatecas, ZAC 98068 Mexico\\
and\\
Institut f\"ur Kernphysik, Universit\"at Mainz, D-55099 Mainz, Germany } }
\maketitle
\begin{abstract}
The key assumption used recently by
Feldmann, Kroll and Stich $\lbrack $Phys.\ Rev.\ {\bf D58}, 114006 
(1998)$\rbrack $ stating that the decay constants $f_\eta $,
and $f_{\eta '}$ of the respective $\eta $ and $\eta '$ mesons
in the quark flavor basis 
follow the pattern of strange and non--strange quarkonia mixing
in their wave functions,
is  reproduced in identifying the non-isotriplet part of the strong neutral 
axial current of the quarks  with the genuine axial hypercharge current
$J_{\mu ,5}^Y =\bar q \gamma_\mu\gamma_5 {Y\over 2}q$, 
where $Y=C+S+B$ is defined by the
Gell-Mann-Nakano-Nishijima relation as the sum of charm $(C)$,
strangeness $(S)$, and baryon $(B)$ quark quantum numbers.
The inequivalence between the octet and hypercharge axial currents
is pointed out.
 \end{abstract}
PACS: 11.40.Ha, 11.30.Hv  \\

Recent investigations on the $\eta (\eta ' )\to 2\gamma $ decays revealed the
necessity for accounting for SU(3)$_F$ violation by introducing two 
different mixing angles between the flavor octet and
singlet states in the wave functions of the $\eta $ and $\eta '$
mesons, respectively, \cite{Leut}. On the contrary,
in Ref.~\cite{Feldmann}, where the octet-singlet basis
was rigorously given up in favor of the quark flavor one,
the decay constants of the strange and non--strange quarkonia
parts of the $\eta (\eta ')$ wave functions,
in turn denoted by $f^s_i$ and $f^q_i$ with $i=\eta , \eta ' $, 
have been successfully calculated in terms of only one meson 
independent mixing angle between the quark flavor states.
The basic starting point in Ref.~\cite{Feldmann} 
is contained in Eqs.~(2.1-2.3) there, where
$f^q_i$, and $f^s_i$ have been assumed to follow the pattern of the 
strange--non-strange quarkonia mixing  according to 
 \begin{equation}
 \left(
 \begin{array}{cc}
 f^q_\eta &f^s_\eta\\
 f_{\eta'}^q&f_{\eta '}^s\end{array}\right)\,  =
 \left(
 \begin{array}{cc}
 \cos\alpha &-\sin\alpha\\
 \sin\alpha&\cos\alpha
 \end{array}\right)
 \left(\begin{array}{cc}
 f^q&0\\
 0&f^s
 \end{array}\right)\, .
\label{mixing}
 \end{equation}
 Here, $f^q$ and $f^s$ in turn denote the decay constants of the
 flavor basis states given by the respective non-strange,
($\eta ^q= {1\over \sqrt{2}}(\bar u u + \bar d d)$) and strange
($\eta^s= \bar s s)$ quarkonia.
An appealing feature of the scheme of Ref.~\cite{Feldmann}
is that the $f^s/f^q$ ratio, in being expressed in terms
of the anomaly matrix elements
$\langle 0|G\widetilde{G}|\eta^q (\eta^s)\rangle $, 
where $G$ is the gluon field tensor, while $\widetilde{G}$ is its dual,
allows to exclude via their Eq.~(2.6) the second mixing angle appearing 
in the original octet-singlet basis of Ref.~\cite{Leut}. 

We here show that both the ansatz in Eq.~(\ref{mixing})
and the appearance of the axial QCD anomaly in the decay
properties of the $\eta $ and $\eta '$ mesons 
find a natural explanation in
identifying the strong current entering the definition of $f_\eta^q$,
 $f_\eta ^s$, $f_{\eta '}^q$, and $f_{\eta '}^s$
with the genuine hypercharge axial current, to be denoted by
 $J_{\mu ,5}^Y$.
 
Recall, that the hypercharge $Y$ in four-flavor space is determined
in a unique way through the  Gell-Mann-Nakano-Nishijima relation: 
\begin{eqnarray}
\hat{Q} =\hat{t}_3 +{{\hat{Y}}\over 2}\, ,&\quad&
{{\hat{Y} }\over 2} = \hat{t}_H +{{\hat{B}}\over 2}\, ,
\label{GLM_NSHI}
\end{eqnarray}
with $2\hat{t}_3=\mbox{diag}(1,-1,0,0)$, 
$2\hat{t}_H =\mbox{diag}\, (0,0,\hat{C},\hat{S})$, and
$\hat{B}=\mbox{diag}\, ({1\over 3},{1\over 3},{1\over 3},{1\over 3})$.
The hypercharge axial current of the fundamental quark quadruplet
$q=$column$(u,d,c,s)$ now reads 
\begin{eqnarray}
J_{\mu ,5}^Y &=& \bar q \gamma_\mu\gamma_5 \hat{t}_H q +
\bar q \gamma_\mu\gamma_5 {{\hat{B}}\over 2} q \, .
\label{hyperch_axcurr}
\end{eqnarray}
To be specific, we here calculate the matrix element of this current between
the vacuum and, say,  the $\eta $ meson state $\mid \eta \rangle $
\begin{eqnarray}
\langle 0\mid J_{\mu ,5}^Y\mid \eta \rangle =f_\eta \, ip_\mu  \, .
\label{ma_el}
\end{eqnarray}
In the following, the wave function of the $\eta $ meson will be
represented in the quark flavor basis where the violation of the OZI
rule has been made explicit according to
\begin{equation}
\mid \eta \rangle  = {1\over \sqrt{2}}(\bar u u +\bar d d)\, \cos \alpha
-\bar s s \, \sin \alpha .
\label{eta_wafu}
\end{equation}
In inserting  Eqs.~(\ref{hyperch_axcurr}) and (\ref{eta_wafu})
into Eq.~(\ref{ma_el})
and in simultaneously assuming that the quarkonium-quark couplings,
denoted by $\kappa_j$, are
diagonal in the flavor $j$, but not necessarily universal in strength, i.e. 
\begin{equation}
\langle 0\mid \bar q_k {1\over 2} \gamma_\mu \gamma_5 q_k
 \mid \bar q_jq_j\rangle
= \delta _{kj}\kappa_j\, ,
\label{Jaffe}
\end{equation}
we reproduce the ansatz of Ref.~\cite{Feldmann} as 
\begin{eqnarray}
\langle 0\mid {1\over 2}(\bar c \gamma_\mu \gamma_5 c 
- {2\over 3}\bar s \gamma_\mu
\gamma_5 s)
&+& {1\over 6}(\bar u\gamma_\mu \gamma_5 u + \bar d\gamma_\mu d +\bar c
\gamma_\mu\gamma_5 c)|{1\over {\sqrt{2}} }(\bar u u +\bar d d)\cos\alpha 
 - \bar ss\,\sin \alpha  \rangle \nonumber\\
= f^s\sin\alpha \, ip_\mu\, &+& f^q\cos\alpha \, ip_\mu \, ,
\nonumber\\
\mbox{with} \quad f^s={2\over 3}\kappa_s\, , &\quad & 
f^q={{\sqrt{2}}\over 3}\kappa_u\, .
\end{eqnarray} 
Here, we assumed $\kappa_u=\kappa_d$ and thereby isospin symmetry
for the light flavors. From the last equation follows that in case
$\kappa_s=\kappa_u=\kappa_d$, the following relation is valid:
\begin{equation}
{f^s\over f^q}=\sqrt{2}.
\end{equation}
The phenomenological value for that ratio found in
Ref.~\cite{Feldmann} reads $f^s/f^q= 1.34\pm 0.06$ and is
compatible with $\sqrt{2}$. 
Nonetheless this situation can not be interpreted
as SU(3)$_F$ symmetry for the simple reason that the hypercharge axial current
is not a genuine SU(3)$_F$ Noether current, an observation already 
reported in \cite{KiPRD}. Indeed, the Gell-Mann-Nakano-Nishijima relation
in Eq.(\ref{GLM_NSHI}) requires hypercharge to act as a generator of a pertinent 
U(4) subgroup, called ${\cal S}_{OZI}$ here, and defined as 
\begin{equation}
{\cal S}_{OZI}=SU(2)_{ud}\otimes 
SU(2)_{cs}\otimes U(1)\, ,
\label{OZI_symm}
\end{equation}
with the two SU(2) groups in turn acting onto the 1st and 2nd quark
generations.
{}From that and Eq.~(\ref{hyperch_axcurr}) one directly reads off that 
$J_{\mu , 5}^Y$ contains inevitably the anomalously divergent axial baryon number
current $J_{\mu , 5}^B = \bar q \gamma_\mu \gamma_5 {B\over 2} q$, 
whose divergency is proportional to $G\widetilde{G}$.
Because of that, the hypercharge axial current is anomalously divergent too.
Indeed, in making use of the abelian axial anomaly relation for a 
single quark species $q_j$,
\begin{equation}
\partial^\mu \bar q_j\gamma_\mu\gamma_5 q_j=2m_j \bar q_j i\gamma_5 q_j
+\alpha_s {{\widetilde{G}G}\over {4\pi }}\, ,
\label{1fl_anom}
\end{equation}
where, $m_j$ is the mass of a quark of flavor $j$, and $\alpha_s$ the 
strong coupling constant, one easily confirms that $J_{\mu ,5}^Y$ 
has no well defined chiral limit
because its divergency contains the U(1)$_A$ QCD anomaly according to

\begin{eqnarray}
\partial^\mu J_{\mu ,5}^Y
= (m_c\bar c i\gamma_5 c - m_s\bar s i\gamma_5 s) &+&
{1\over 3} (m_u\bar u i\gamma_5 u + m_d \bar d i\gamma_5 d +
m_c\bar c i\gamma_5 c + m_s\bar s i\gamma_5 s)
+\alpha_c {  { \widetilde{G}G }\over {6\pi } }\,.
\label{4fl_anom}
\end{eqnarray}
In this way the contribution of the gluons enters the
calculation of the $\eta  $ (as well as  $\eta ')$ decay properties
and accounts for the successful data interpretation presented in 
\cite{Feldmann}.
Note that despite the circumstance, that in
the truncated three flavor space the hypercharge
happens by chance to coincide numerically
with the octet SU(3)$_F$ generator, the octet and hypercharge axial
currents are quite different observables.
Indeed, in contrast to the hypercharge axial
current the octet one, 
$J_{\mu ,5}^8 =\bar q_3\gamma_\mu \gamma_5 \frac{\lambda^8}{2} q_3$
with $\sqrt{3}\lambda_8 = 2t_3 +4t_3^U$,
and $t_3^U=\mbox{diag}(0,1,-1)$, is conserved in
the chiral limit of vanishing quark masses of the quark triplet
$q_3=$column$(u,d,s)$:
\begin{eqnarray}
\lim_{m_j\to 0} \partial^\mu J_{\mu ,5}^8 &=&
\lim_{m_j\to 0} \frac{1}{\sqrt{3}}(m_u\bar u i\gamma_5u
-m_d\bar d i\gamma_5 d)
 +\frac{2}{\sqrt{3}}(m_d\bar d i\gamma_5d - m_s\bar s i\gamma_5 s)  \to 0\, ,
\label{octet_ax}
\end{eqnarray}
as the anomaly contributions in the isospin and U-spin pieces cancel out,
separately.
The nucleon matrix element of the hypercharge axial current determines
the $\eta N$ coupling constant and has been proven in \cite{KiPRD}
to be also well compatible with data.


\end{document}